\documentclass[conference]{IEEEtran}
\IEEEoverridecommandlockouts
\usepackage[utf8]{inputenc}
\usepackage{xcolor}
\usepackage{graphicx}
\usepackage{amsmath}
\usepackage[version=4]{mhchem}
\usepackage{siunitx}
\usepackage{float}
\usepackage{subcaption}
\usepackage{multirow}
\usepackage{longtable,tabularx}
\usepackage{physics}
\usepackage{booktabs}
\usepackage{nomencl}
\usepackage{mathtools, cuted}
\usepackage{mathtools}
\usepackage{calc}
\newcommand{\DNN}{Deep Neural Networks }
\newcommand{\INN}{Invertible Neural Networks }
\newcommand{\lrems}{LREM }

\title{Efficient Inverse Design of 2D Elastic Metamaterial Systems Using Invertible Neural Networks}

\author{\IEEEauthorblockN{Manaswin Oddiraju \IEEEauthorrefmark{1}, 
Amir Behjat\IEEEauthorrefmark{2}, 
Mostafa Nouh\IEEEauthorrefmark{3}, 
Souma Chowdhury\IEEEauthorrefmark{4}}
\IEEEauthorblockA{\textit{Department of Mechanical \& Aerospace Engineering} \\
\textit{University at Buffalo}\\
Buffalo, NY, 14260\\
Email:\IEEEauthorrefmark{1}moddiraj@buffalo.edu, 
\IEEEauthorrefmark{2}amirbehj@buffalo.edu,
\IEEEauthorrefmark{3}mnouh@buffalo.edu,
\IEEEauthorrefmark{4}soumacho@buffalo.edu}}

\begin{document}

\maketitle  

\begin{abstract}
Locally resonant elastic metamaterials (LREM) can be designed, by optimizing the geometry of the constituent self-repeating unit cells, to potentially damp out vibration in selected frequency ranges, thus yielding desired bandgaps. However, it remains challenging to quickly arrive at unit cell designs that satisfy any requested bandgap specifications within a given global frequency range. This paper develops a computationally efficient framework for (fast) inverse design of LREM, by integrating a new type of machine learning models called invertible neural networks or INN. An INN can be trained to predict the bandgap bounds as a function of the unit cell design, and interestingly at the same time it learns to predict the unit cell design given a bandgap, when executed in reverse. In our case the unit cells are represented in terms of the width's of the outer matrix and middle soft filler layer of the unit cell. Training data on the frequency response of the unit cell is provided by Bloch dispersion analyses. The trained INN is used to instantaneously retrieve feasible (or near feasible) inverse designs given a specified bandgap constraint, which is then used to initialize a forward constrained optimization (based on sequential quadratic programming) to find the bandgap satisfying unit cell with minimum mass. Case studies show favorable performance of this approach, in terms of the bandgap characteristics and minimized mass, when compared to the median scenario over ten randomly initialized optimizations for the same specified bandgaps. Further analysis using FEA verify the bandgap performance of a finite structure comprised of $8\times 8$ arrangement of the unit cells obtained with INN-accelerated inverse design. 
\end{abstract}
\makenomenclature

\printnomenclature
\section{Introduction}
Metamaterials can be thought of as compound structures that consist of self-repeating smaller substructures referred to as ``unit cells'' \cite{fok2008acoustic,al2017formation}. Such a composition enables them to have customizable physical properties which are otherwise not naturally feasible \cite{fok2008acoustic,chronopoulos2017enhanced}. As such, they have become increasingly popular for 
potential applications in acoustics \cite{sui2015lightweight,mamaghani2016vibration}, photonics \cite{liu2008three,gansel2009gold}, sensing and other end uses \cite{chen2012metamaterials,albabaa2019emergence,nouh2019control}. 
Owing to the periodic nature of traditional metamaterial systems, the standard approach is to use an individual cell to predict the dynamic response characteristics \cite{nouh2015wave} of the overall structure \cite{ge2007ga,nouh2014vibration}, thereby reducing the design problem to optimizing the design of the unit cell. 
In the case of mechanical metamaterials, the dynamic response of these structures contain ``stop bands'' and ``passbands''. Vibrations with frequencies in these passbands are propagated while the frequencies lying in the stopbands are damped. The design of such metamaterial systems with specified desired properties has traditionally been a computationally challenging problem as it requires the evaluation of expensive numerical simulations \cite{wilt2020accelerating} to search through (often) highly constrained design spaces. Moreover, larger complex system level design efforts call for fast on-demand retrieval of unit cells (satisfying specified properties) with which portion of system structure will be made. This calls for significantly improved computational efficiency of the search process for optimum retrieval of self-repeating unit cells. Machine learning (ML) methods provide avenues to fill this important gap in the metamaterial design paradigm, examples of which include the use of Artificial Neural Networks (ANN) \cite{suzuki2013artificial}, Generative Adversarial Networks(GAN) \cite{goodfellow2014generative} and Gaussian Process architectures \cite{gramacy2020surrogates}. In this paper, we explore the use of an emerging type of ML model, namely invertible neural networks or INN \cite{inversenets,metamaterial_IDETC_2020}, to speed up the process of designing 2D Locally Resonant Elastic Metamaterials (LREM); in this context, the INN uniquely enables instantaneous retrieval of unit cells satisfying specified bandgap properties. 
%
The remainder of this section is ordered as follows: we first introduce the concept of LREM, before giving a brief literature review of ML methods used in the design of LREM and related metamaterials, and converging on the objectives and outline of this paper.

\subsection{Locally Resonant Elastic Metamaterials}
\label{sec:meta_material}
\begin{figure*}[t]
\centering
     \begin{subfigure}{0.3\linewidth}
        \includegraphics[width=\linewidth]{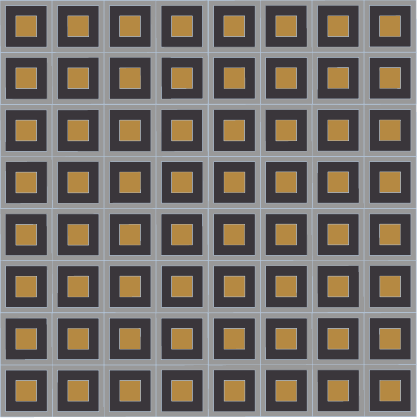} 
        \caption{}
        \label{fig:2dplate}
     \end{subfigure}
     \hspace{1cm}
     \begin{subfigure}{0.3\linewidth}
         \includegraphics[width=\linewidth]{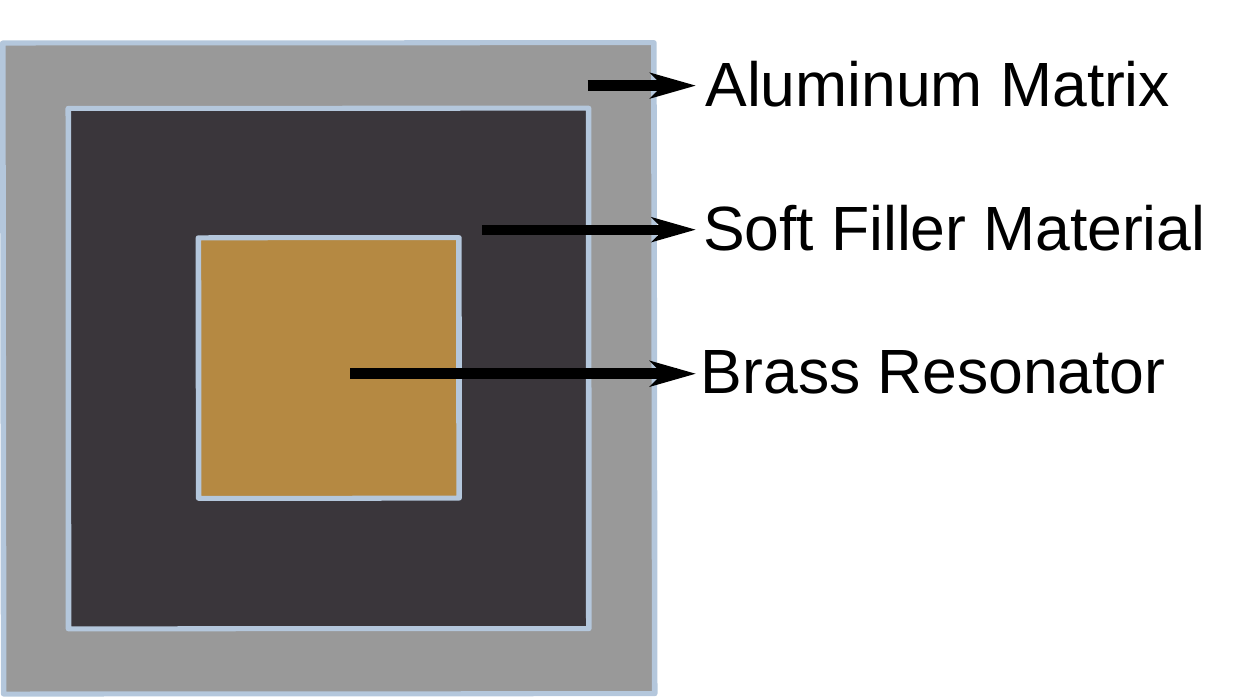} 
         \caption{}
     \end{subfigure}
    \caption{2D Metamaterial system: a) Metamaterial plate; b) Hard-Soft-Hard unit cell}
    \label{fig:2D_MM}
\end{figure*}

\nomenclature{LREM}{Locally Resonant Elastic Metamaterials}
\nomenclature{INN}{\INN}
\nomenclature{DNN}{\DNN}
Locally Resonant Elastic Metamaterials or LREM are a class of periodic structures comprised of a host medium and an array of internal mechanical resonators \cite{nouh2014vibration}. Such a combination allows for dynamic interactions between the different materials and gives rise to interesting dynamic phenomenon called "Bandgaps"  \cite{sui2015lightweight,mamaghani2016vibration}. An example of one such configuration in two-dimensional space is the hard-soft-hard plate shown in Fig.~\ref{fig:2D_MM}, in which a self-repeating (or periodic) arrangement of square (or rectangular) cells consist of an outer steel matrix containing a soft filler layer and a rigid concentric mass. The filler-mass combination provides the local resonance mechanism (LR) necessary to induce the LR band gaps \cite{al2017formation}. Research efforts have seen the application of such elastic metamaterials in areas ranging from noise control \cite{ang2017broadband,gao2018composite} and mitigation in engine mounts, turbines and automotive parts \cite{jung2019realisation}, to new and unprecedented avenues where such metamaterials open up the possibility of logic gates, topological insulators, diodes, and switch-like functions in mechanical systems analogous to electronic circuits \cite{bilal2017bistable}. Since commonly used damping materials (e.g., viscoelastic and rubber-like polymers) usually have mediocre load-bearing capability \cite{lakes2009viscoelastic}, elastic metamaterials presents a lucrative alternative for pseudo-damping mechanisms that achieve targeted attenuation behavior over frequency ranges of interest. 

Due to the reasons mentioned earlier, notable fraction of emerging metamaterial design methods exploit data-driven techniques. For instance,  
Bostanabad et al.  \cite{bostanabad2019globally} used Gaussian Process for big data analysis in a process called Globally Approximate Gaussian Process (GP) and applied it to design metamaterials. 
Hpwever, with approach, a separate optimization is required every time a user desires a different specified property. Another recent approach uses a Bayesian Network Classifier \cite{matthews2016hierarchical,morris2018design}, which significantly reduces computational effort and provisions model error measures. However, it is also not directly amenable to efficient inverse design or on-demand retrieval of unit cells, given desired properties. 
In another work, Ma et al. 
\cite{ma2019probabilistic} used a variational autoencoder-based generative model to design photonic metamaterials. Such approaches \cite{liu2018generative} often use Generative Adversarial Networks (GAN), which in turn rely on large random data sets to train. Another approach uses an adaptive neural network to achieve similar objectives \cite{chen2019smart}. 

The ML models in the above-stated approaches do not guarantee a computable inverse for retrieving the unit cell design, given specified properties. INNs instead have a unique architecture that allows them to preserve an analytically computable Jacobian, thereby guaranteeing an exact inverse can be retrieved by feeding the output and running the INN backwards. In our previous work  \cite{metamaterial_IDETC_2020}, we exploited this concept of INN to design 1D phononic structures with application to vibration suppression in drill strings with annular inserts. There, INNs were trained to model the discretized frequency response of periodic and aperiodic structures. These
INN models were used to instantaneously retrieve
approximate inverse designs with new user-desired frequency responses, which are then used to initialize a constrained
optimization process to find more accurate inverse designs, resulting in significantly improved outcomes compared to randomly initialized design optimizations. Here, we extend the applicability of this inverse design concept to a significantly more complex 2D problem, where the property of interest is low frequency bandgaps in LREM. This further development requires more complex high-fidelity analyses for generating training samples and validating final designs, and re-imagining how to encode the output space of the INN to model bandgaps, i.e., if they exist within the specified global frequency range. 
\subsection{Research Objectives of this Paper}
With the given background and need for fast retrieval of inverse designs, the overall objectives of this paper can be summarized as:
\begin{enumerate}
    \item Identify a representation of bandgaps to be treated as an output that INNs, and for comparison standard deep neural network (DNNs), can predict as a function of unit cell design.
    \item Develop an inverse design framework for 2D LREMs, by integrating dispersion analysis for generating high-fidelity samples of frequency response, INN and DNN models respectively for inverse retrieval and surrogate based optimization initialized thereof, and optimization solvers. 
    \item Use the inverse design framework to find 2D LREMs that satisfy different user-specified bandgap constraints while having the minimum mass, and compare its computational efficiency and resulting mass and bandgap with those of randomly-initialized design optimizations. 
    \item Develop a finite element analysis model to be used for verifying the existence of the predicted bandgaps in a finite LREM system comprising an $8\times 8$ arrangement of designed unit cells.
\end{enumerate}

The remainder of this paper is ordered as follows: In Section \ref{sec:framework}, we first present an overview of our design framework before covering the individual components in detail. Section \ref{sec:problem_def} contains the details of our inverse design problem, including our design variables, INN modelling parameters and optimization settings. In Section \ref{sec:Results} we present the results and its discussion, and thereafter ending with concluding remarks in Section \ref{sec:Conclusion}

\section{INN Based Inverse Design Framework}
\label{sec:framework}
\begin{figure*}
    \begin{subfigure}{\linewidth}
        \centering
        \includegraphics[width=0.7\linewidth]{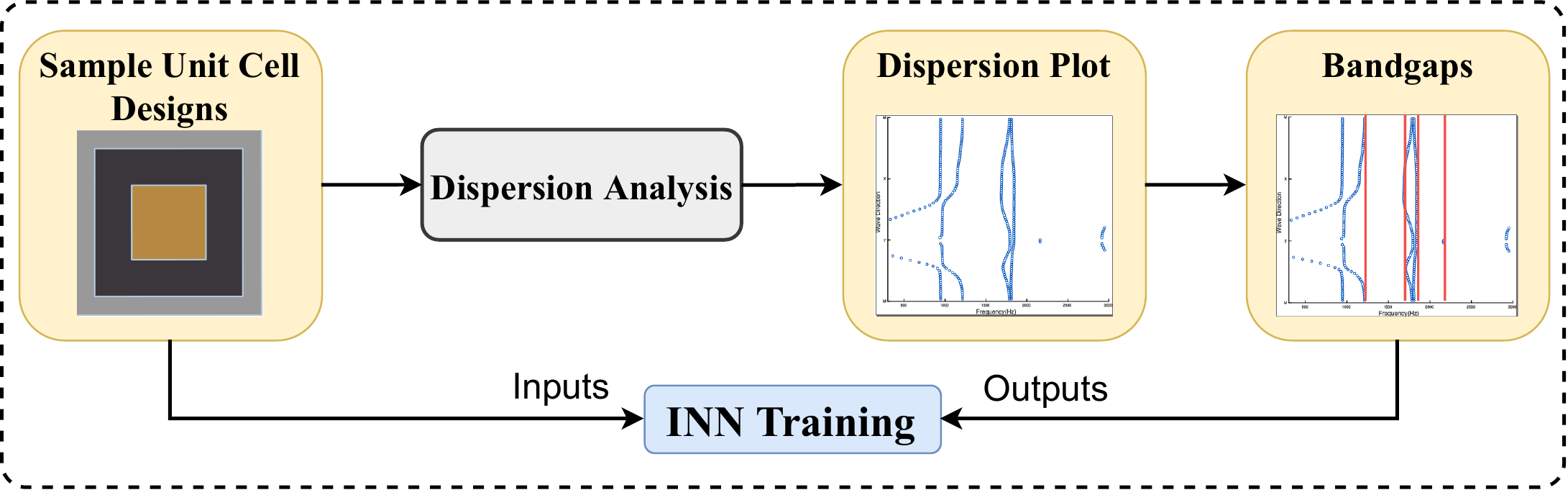}
        \caption{}
    \end{subfigure}
    
    \begin{subfigure}{\linewidth}
        \centering
        \includegraphics[width=\linewidth]{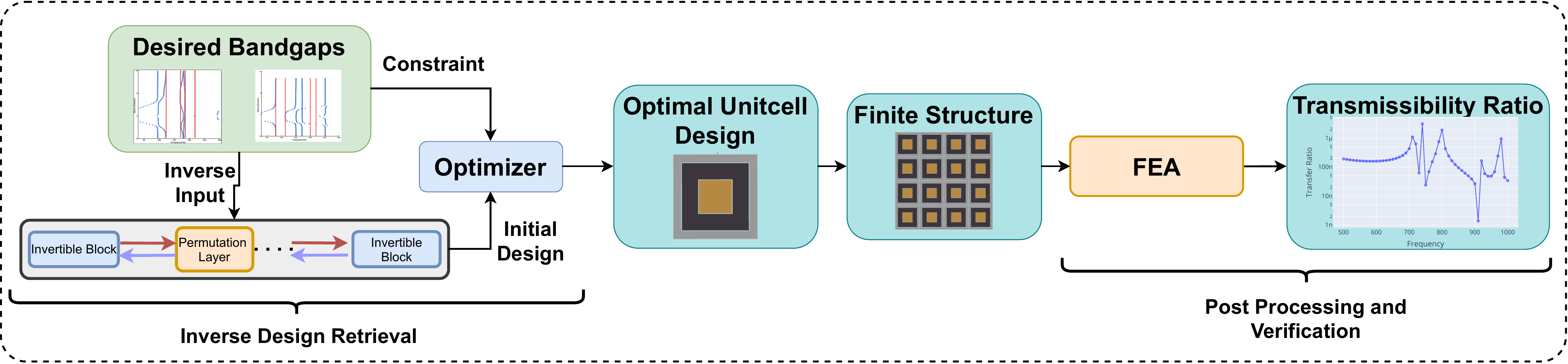}
        \caption{}
    \end{subfigure}
    \caption{INN accelerated inverse design framework for 2D LREM: a) INN training; b) Inverse design retrieval and INN initialized optimization}
\label{fig:framework}
\end{figure*}

Figure \ref{fig:framework} shows the overall framework developed in this paper. First we start with training the INN on samples generated by dispersion analysis and then using the trained INN in inverse direction to generate an initial design for optimization. Our hypothesis is that the INN will generate constraint satisfying designs, i.e designs which have bandgaps that satisfy the user requirements. Starting at a feasible design (or even close to a feasible design) will reduce the computational cost of the forward optimization process and/or enhance the quality of the design found with a given function evaluation investment. It is important to note that, there is no guarantee that constraint satisfying inverse designs do exist for any given user specifications of desired bandgaps. Below, we describe the critical parts of the framework, namely Bloch dispersion analysis used for generating training and testing samples, the INN architecture used for inverse retrieval of design, the optimization approach used, and the finite element model developed for verification of the optimized designs.


\subsection{Bloch Dispersion Analysis}
\label{sec:dispersion}
The dispersion analysis is a useful tool that helps us in determining the frequencies in which a metamaterial can have wave propagation and therefore by extension, can also help us in finding the frequencies where there is no wave propagation or ``Bandgaps". The dispersion analysis is done by first discretizing the the unit cell into a 2D Finite element grid with 2D Mindlin plate elements and then assembling the global stiffness and mass matrices and also the displacement vectors. After that, a set of periodic boundary conditions also known as the ``Bloch-Floquet" \cite{hakoda2018using} boundary conditions as shown in Eq. [\ref{eq:periodic_bc}] are applied.

\nomenclature{$q$}{Displacement at node}
\nomenclature{$\tilde{k}_{x}$}{Wave number in x-direction}
\nomenclature{$\tilde{k}_{y}$}{Wave number in y-direction}

\begin{figure}
    \centering
    \includegraphics[width=\linewidth]{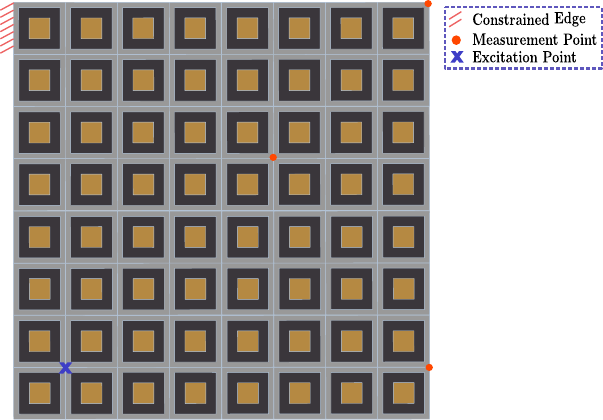}
    \captionof{figure}{FEA boundary conditions}
    \label{fig:fea_bc}
\end{figure}

\begin{equation}
\footnotesize
\label{eq:periodic_bc}
\begin{aligned}
q_{o}=& T \hat{q}\\
\left[\begin{array}{ccccccccc}
q_{I} &
q_{B} &
q_{T} &
q_{L} &
q_{R} &
q_{L B} & 
q_{R B} &
q_{L T} &
q_{R T}
\end{array}\right]^{\rm{T}} =& T \hat{q}\\
\left[\begin{array}{cccc}
I & 0 & 0 & 0 \\
0 & I & 0 & 0 \\
0 & I e^{\tilde{k}_{y} i} & 0 & 0 \\
0 & 0 & I & 0 \\
0 & 0 & I e^{\tilde{k}_{x} i} & 0 \\
0 & 0 & 0 & I \\
0 & 0 & 0 & I e^{\tilde{k}_{x} i} \\
0 & 0 & 0 & I e^{\tilde{k}_{y} i} \\
0 & 0 & 0 & I e^{\left(\tilde{k}_{x}+\tilde{k}_{y}\right) i}
\end{array}\right]\left[\begin{array}{c}
q_{I} \\
q_{B} \\
q_{L} \\
q_{L B}
\end{array}\right] =& T \hat{q}
\end{aligned}
\end{equation}
and where, q is the nodal displacement and the subscripts $I,B,T,L,R,LB,RB,LT,RT$ indicate the positions of nodes at interior, bottom, top, left, right, left-bottom, right-bottom, left-top and right-top respectively. $k_x,k_y$ are the wave numbers in x and y directions.
Then, $\omega$ is obtained by solving the following eigen-problem:
\begin{equation}
   (\hat{K_0}-\omega^2\hat{M_0})\hat{q} = 0 \text{  where, } 
\end{equation}
Where, $\hat{K_0} = \bar{T}^T K_0 T$ and $\hat{M_0}= \bar{T}^T M_0 T$. 

By varying $(k_x,k_y)$ along $(\pi,\pi) \xrightarrow{} (0,0) \xrightarrow{} (\pi,0) \xrightarrow{} (\pi,\pi)$ (called the ``Irreducible Brillouin Zone" \cite{craster2012dangers}) we get a vector of frequencies($\boldsymbol{\Omega} = [\omega_1,\omega_2,\dots]$) at which wave propagation occurs in the metamaterial. From these frequencies, the bandgaps are filtered out as follows: 
\nomenclature{$\boldsymbol{\Omega}$}{Set containing the frequencies in which wave propogation occurs in the metamaterial}
\nomenclature{$\boldsymbol{\Omega}_{\rm{BG}}$}{Set containing the bandgap frequencies}
\nomenclature{$\omega_i$}{$i^{\rm{th}}$ frequency in set $\boldsymbol{\Omega}$ when sorted in ascending order }
\begin{equation}
\begin{aligned}
      \boldsymbol{\Omega}_{\rm{BG}} &= \boldsymbol{\Omega}_{\rm{A}} \cap \boldsymbol{\Omega}_{\rm{B}} \text{ where,} \\
      [\omega_i,\omega_{i+1}] \in \boldsymbol{\Omega}_{\rm{A}} &\ni  \omega_{i+1}-\omega_i > 20 \text{Hz}\& i \in [1,\boldsymbol{n}(\boldsymbol{\Omega})] \\ 
      [\omega_i,\omega_{i+1}] \in \boldsymbol{\Omega}_{\rm{B}} &\ni  \delta(\textbf{C}(\omega_i,\boldsymbol{\Omega}),\textbf{C}(\omega_{i+1},\boldsymbol{\Omega})) = 15 
\end{aligned}
\end{equation}
In the above equations, the operator \textbf{C}$(\omega_i,\boldsymbol{\Omega})$ represents the number of times $\omega_i$ occurs in the set $\boldsymbol{\Omega}$; and 
\begin{equation}
    \delta(\omega_i,\omega_{i+1}) = \begin{cases}
15 &\text{if $\omega_i \geq 15$ or $\omega_{i+1} \geq 15$ }\\
 0 &\text{otherwise}
\end{cases} 
\end{equation}

\subsection{Invertible Neural Networks}
\INN \cite{inversenets} are a type of neural networks which in addition to predicting an output given the input, can also predict the input given an output when run in reverse. The unique advantage of the INN is that they can learn both the forward and inverse mappings at the same time.  
A handful of notable INN architectures have been reported in recent years \cite{mccann2017convolutional}, with preliminary applications to image reconstruction, parameter estimation, and generative flow modeling. We chose to implement the Invertible Network architecture proposed in \cite{inversenets}, since it offers an exactly computable inverse. This architecture ensures invertibility by preserving the non-singular property of the Jacobian of the activations across a layer. We used a similar architecture of the INN as in our previous work \cite{metamaterial_IDETC_2020}. The following equations show the working principle of the INN: 
\begin{equation}
\begin{aligned}
     & v_1 = u_1 \odot exp(s_2(u_2))+t_2(u_2) , \\ & v_2 = u_2 \odot exp(s_1(v_1)) + t_1(v_1) 
    \end{aligned}
\end{equation}
Here, $u$ and $v$ are the inputs and outputs of the Invertible Network and the subscripts 1 and 2 indicate their first and second halves, respectively. $s_1$, $s_2$, $t_1$, and $t_2$ are neural networks with fully connected layers with Leaky ReLU activation functions, and $\odot$ represents element-wise multiplication. 

The above expressions can be easily inverted if the output $v = [v_1 \text{, } v_2]$ is known
\begin{equation}
\begin{aligned}
     & u_2 = (v_2 -t_1(v_1)) \odot exp(-s_1(v_1)) ,\\ & u_1 = (v_1-t_2(u_2))\odot exp(-s_2(u_2))
    \end{aligned}
\end{equation}
Further details and illustration of the INN architecture used here are given in \ref{appendix:INN}. 

\subsection{Optimization Approach: Sequential Quadratic Programming (SQP)}
In this paper we use a standard gradient based solver, the Sequential Quadratic Programming (SQP) \cite{kelley1999iterative}, to perform the constrained forward optimization for finding the unit cell designs that satisfy the user-specified bandgap constraints and minimizes the system mass. SQP is implemented using MATLAB's built-in optimization toolbox, Although we knew from our initial numerical experiments that the problem tends to be multi-modal, we deliberately chose SQP so as to eliminate any additional stochastic effects (otherwise unavoidable with metaheuristic solvers such as genetic algorithms or particle swarm optimization) when comparing the results of INN-initialized optimizations with a set of randomly initialized optimizations. 


\subsection{Finite Element Analysis (Postprocessing and Verification)}
\label{sec:fe_description}
The advantage of using the dispersion analysis is that it is capable of determining the bandgaps definitively. However, the boundary conditions used in the dispersion analysis assume that the unitcell is part of an \textit{infinite} structure with infinite number of unitcells along each direction. Therefore, the bandgaps that are predicted by the dispersion analysis may not be present in an actual \textit{finite structure} composed of a finite number of cells as one would expect in a practical system. Hence, to verify the realization of the (dispersion analysis) predicted bandgap in a finite structure, we perform a harmonic FEA analysis and plot the transmissibility ratio of the finite structure (in this case a plate with $8\times 8$ grid of unitcells) across the frequency range of interest. The transmissibility ratio is computed as the ratio between the observed displacement at a predetermined node to the magnitude of the input force. Table \ref{tab:fea_params} gives the details of the FEA tools and setup used and Section \ref{sec:fe_parameters} provides the detailed parameters of our analysis.
\begin{table}[]
	\caption{Harmonic analysis parameters}
		\begin{tabularx}{\linewidth}{>{\centering}X >{\centering}X}
            \toprule
            Analysis Parameter & Value  \tabularnewline
            \midrule
              Analysis Type & Harmonic Analysis \tabularnewline
              FEA Package & Code Aster   \tabularnewline
              Element Type & 6 node triangular element \tabularnewline
              Time Stepping & Newmark's Method \tabularnewline
            \bottomrule
            \end{tabularx}
		\label{tab:fea_params}
\end{table}
\begin{table}[]
	\caption{Design variable bounds}
	\centering
		\begin{tabularx}{\linewidth}{>{\centering}X >{\centering}X >{\centering}X}
            \toprule
            Variable & Bounds  \tabularnewline
            \midrule
              $x_m$ & [1,3] mm  \tabularnewline
              $y_m$ & [1,5] mm \tabularnewline
              $x_f$ & [1,3] mm  \tabularnewline
              $y_f$ & [1,5] mm  \tabularnewline
            \bottomrule
            \end{tabularx}
		\label{tab:var_bounds}
\end{table}
\begin{table}[]
    \centering
    \caption{Neural network settings}
    \begin{tabularx}{\linewidth}{ >{\centering}m{2.6cm} >{\centering}X >{\centering}X }
    \toprule[0.12em]
    \textbf{Parameter} & \textbf{INN} & \textbf{DNN} \tabularnewline
    \midrule[0.12em]
    Input Size    &  \multicolumn{2}{c }{4} \tabularnewline

    Output Size    & \multicolumn{2}{c }{4}\tabularnewline
    Nodes per Layer & 100 & 150 \tabularnewline

    Hidden Layers    &  4 & 7  \tabularnewline

    Invertible Blocks   & 10 & - \tabularnewline

Type & \multicolumn{2}{c }{Fully Connected} \tabularnewline

Activation Func. & \multicolumn{2}{c }{Leaky ReLU} \tabularnewline

Learning Rate    &  \multicolumn{2}{c }{$10^{-4}$}\tabularnewline

    Optimizer   & \multicolumn{2}{c }{ADAM} \tabularnewline
    \bottomrule[0.12em]
    \end{tabularx}
    \label{tab:network_param}
\end{table}

\section{Problem Definition}
\label{sec:problem_def}
For this paper, we chose to design a unitcell that will ultimately be used in a periodic 2D metamaterial plate containing a grid of $8\times 8$ unitcells, as shown in Fig. \ref{fig:2dplate}. For the optimization, the design variables include the width of the hard matrix and soft filler layers in $x$ and $y$ directions, as shown in Fig. \ref{fig:Design Variables}. The optimization is aimed at reducing the mass of the unitcell and by extension the 2D plate, while satisfying a specified bandgap constraint (i.e., the existence of a bandgap in a specified frequency range). The rest of this section presents the implementation details of the dispersion analysis, optimization formulation and inverse modeling.

\subsection{Dispersion and Harmonic Analysis Parameters}
\label{sec:fe_parameters}
For the purpose of this paper, we limit the frequency range under consideration to 0 Hz -- 2000 Hz. 
Material parameters for different parts of the structure are given in Table \ref{fig:2D_MM}. 
\begin{table*}[]
\begin{minipage}{0.4\linewidth}
    \centering
    \includegraphics[width=0.7\linewidth]{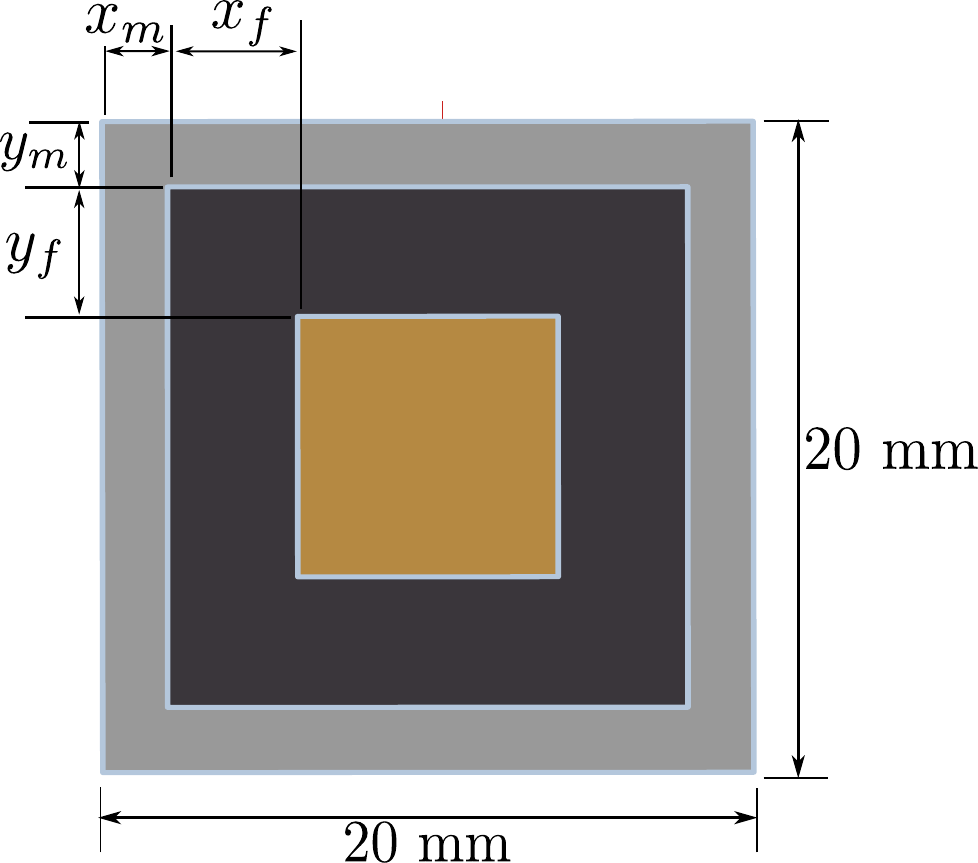}
    \captionof{figure}{Design variables}
    \label{fig:Design Variables}
\end{minipage}\hfill
\begin{minipage}{0.58\linewidth}
	\centering
	\captionof{table}{Structural parameters}
		\begin{tabularx}{\linewidth}{>{\centering}X >{\centering}X >{\centering}X  >{\centering}X}
            \toprule
                Property & Matrix & Filler & Resonator \tabularnewline
                \midrule 
                Elastic Modulus: & 68.9 GPa  & 14.0 GPa & 106 GPa\tabularnewline
                
                Poisson’s Ratio :& 0.33& 0.43 & 0.32\tabularnewline
                
                Density ($\rho$):& 2700 Kg$/\rm{m}^{3}$ & 1565 Kg$/\rm{m}^{3}$ & 8500 Kg$/\rm{m}^{3}$\tabularnewline
                
                Thickness ($t$):& 2 mm &2 mm &3 mm \tabularnewline
            \bottomrule
            \end{tabularx}
		\label{table:struct_props}
\end{minipage}
\end{table*}

Once the optimal design is found, it is verified by comparing its transmissbility ratio measured at the points shown in Fig. \ref{fig:fea_bc} using FEA modeling as described in Section \ref{sec:fe_description}. Here, Fig. \ref{fig:fea_bc} shows the applied boundary conditions and the loading of the metamaterial plate.
\subsection{Optimization formulation}
\label{sec:optimization}
The optimization problem for minimizing the mass of the 2D system subject to the specified bandgap constraints can be formulated as given in Eq. \ref{eq:objfunc}.
\begin{strip}
\centering
\begin{equation}
\label{eq:objfunc}
    \begin{aligned}
  \underset{X}{\min} &: \hspace{10pt} M(X) =  t_{m}\rho_{\rm{M}}(a^2-(a-x_m)(a-y_m))) + t_{f}\rho_{\rm{R}}(a-x_m-x_f)(a-y_m-y_f) + \\ & \hspace{21mm} t_{r}\rho_{\rm{F}}((a-x_m)(a-y_m)-(a-x_m-x_f)(a-y_m-y_f)) \\
        {\rm{s.t.}}& \hspace{10pt}
        g_1(X,\omega_a,\omega_b) : \boldsymbol{n}(\rm{W_{IN}}) - \boldsymbol{n}(\rm{W_{BG}} \cap \rm{W_{IN}}) \leq 0 & \\ 
        {\rm{where,}}&\hspace{30pt}~~  X = \left[x_m,x_f,y_m,y_f\right]
\end{aligned}
\end{equation}
\end{strip}
Here, $a = 20$ mm is the side length of the unitcell as shown in Fig. \ref{fig:Design Variables} and $t_m,~t_f$ and $t_r$ are the thickness of the matrix, filler and resonator respectively, the values of which are given in Table \ref{table:struct_props}. Here, $\rm{W_{IN}}$ and $\rm{W_{BG}}$ are the sets containing all the integer frequency values between $\omega_a$ and $\omega_b$ (the start and end points of the desired bandgap) and between $\boldsymbol{\omega}_1$ and $\boldsymbol{\omega}_2$ ($\boldsymbol{\omega}_1, \boldsymbol{\omega}_2 \in \boldsymbol{\Omega}_{\rm{BG}}$) respectively. And $\boldsymbol{n}$ is the set cardinality operator.

\subsection{INN and DNN Modeling}
\label{sec:inn-params}
We train both an INN and a DNN to map the four design parameters of the unitcell (as shown in Fig. \ref{fig:Design Variables}) to the start and end frequencies of the bandgap (i.e., $\omega_1$ and $\omega_2$), as shown in Eq.~(\ref{eq:inn_modelling})). Note that for the INN the outputs are repeated in order to satisfy the requirement that the input and output layers of the INN have equal number of nodes. The DNN is first used to benchmark the forward prediction accuracy of the INN, which is known to slightly compromise on expressability otherwise achievable with a DNN that is not restricted to provide a computable inverse. For designs which do not show a bandgap in the frequency range under consideration, arbitrary frequency values greater than 2000 Hz (which is the upper bound of frequency range of interest) were assigned. Table \ref{tab:network_param} contains the implementation parameters of both the DNN and the INN. The DNN is later also used as a surrogate model in the constrained forward optimization process to explore how it might further speed up that process.  


\section{Results}
\label{sec:Results}
\nomenclature{RMSE}{Root Mean Squared Error}
We tested our framework by comparing the modeling performance of the INN to a traditional DNN, and later comparing the results of the optimizations initialized with INN-retrieved design to that of randomly initialized optimizations. In both cases, two sets of optimizations are performed, one using the high fidelity Bloch dispersion analysis (BDA) model and the other using the DNN as a surrogate model to predict the bandgaps of candidate designs. The following subsections presents and discusses the results that we obtained.

\subsection{Modeling performance of the INN}
\label{subsec:INN_perf}
\begin{figure}
    \centering
    \includegraphics[width =\linewidth]{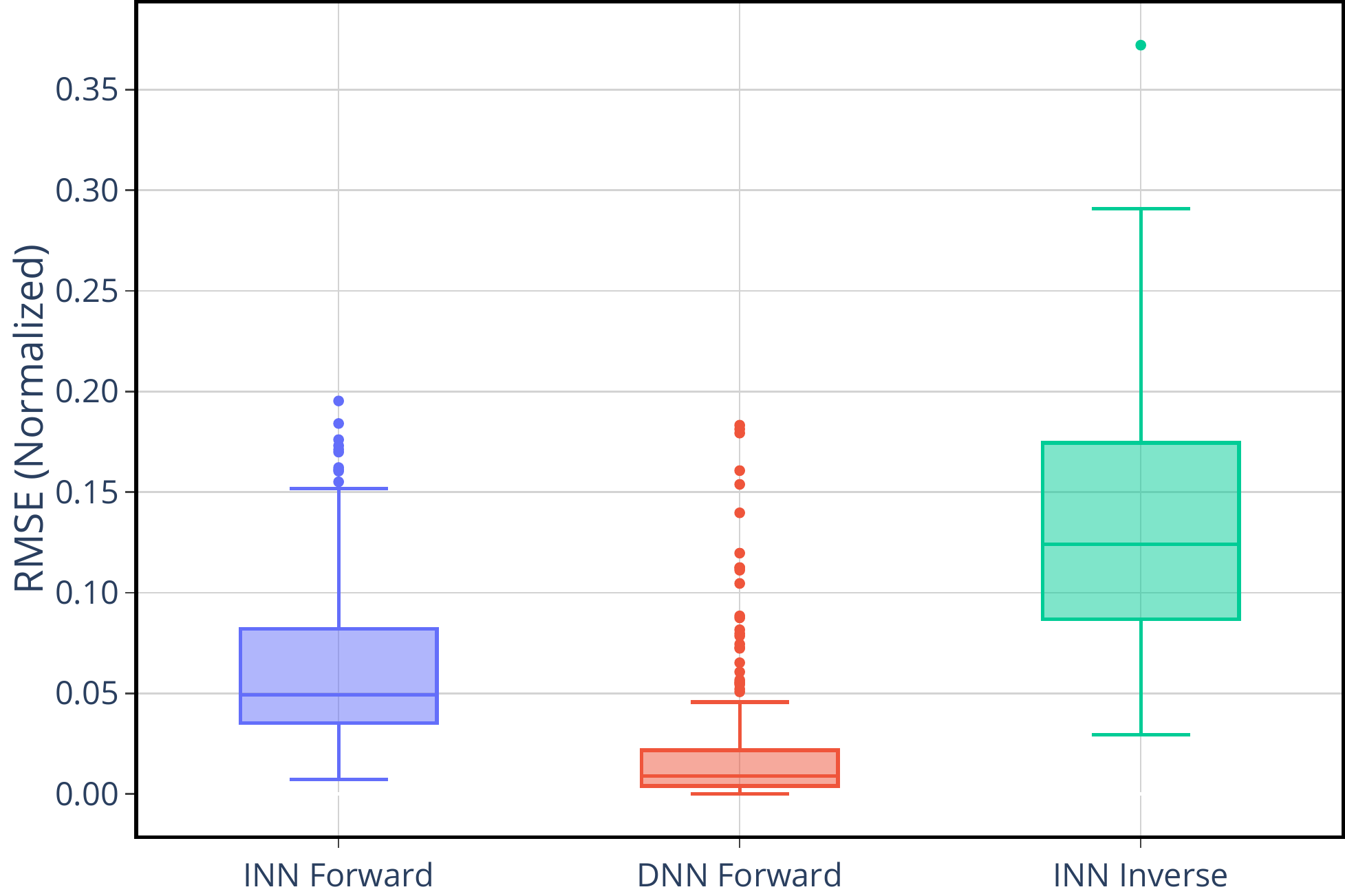}
    \captionof{figure}{Normalized RMSE over 200 unseen test samples}
    \label{fig:rmse}
\end{figure}

We train an INN and a DNN on normalized samples to predict the bandgaps given by the unitcell design. A training set of 1800 samples generated by the dispersion analysis are used. To compare the modeling accuracy of the two models, their forward prediction performance over a set of 200 unseen test samples is compared. We also estimate the prediction performance of the INN in the inverse direction over these test samples. The root mean squared error (RMSE) is used to analyze performance, which for a given sample can be expressed as: \\
\begin{align}
\centering
&\text{Forward}: &  \mathrm{MSE}_\mathrm{Fwd}(X_{i})= \frac{1}{4} \sum_{j=1}^{4}\left(\hat{\boldsymbol{\omega}}_{i j}-\boldsymbol{\omega}_{i j}\right)^{2}\\
&\text{Inverse:} &  \mathrm{MSE}_\mathrm{Inv}(Y_{i})= \frac{1}{4} \sum_{j=1}^{4}\left(\hat{X}_{i j}-X_{i j}\right)^{2}
\label{eq:test-normalized MSE} \vspace{-0.6cm}
\end{align}
where $X_{i}$ is the $i^{\rm{th}}$ sample in the test dataset and $\hat{\boldsymbol{\omega}}_{i j}$ and $\boldsymbol{\omega}_{i j}$ are respectively the forward predictions and the ground truth of the $i^{\rm{th}}$ sample, all data being pre-normalized in this case. Here, $\boldsymbol{\omega}_1 = \boldsymbol{\omega}_3$ and $\boldsymbol{\omega}_2 = \boldsymbol{\omega}_4$. $\hat{X}_{i j}$ is the inverse prediction of the INN. 
Figure \ref{fig:rmse} gives boxplots of the distribution of the testing RMSE per sample. From the figure, it can be seen that the while as expected the DNN is more accurate in forward predictions, the accuracy of INNs at around 5\% median RMSE is quite favorable. The accuracy of the INN in inverse prediction is slightly lower, at a median RMSE of around 12\%. 
In the data obtained from the BDA, the samples which do not show a bandgap between 0-2000 Hz were assigned arbitrary values albeit in a specific range, these data points act as outliers in the data and can lead to increased modeling errors. Such outliers can also affect the modeling performance of the INN more than a traditional DNN as it is highly likely that their presence has a more noticeable impact on the invertibility of the mapping itself. Even though we tried to minimize the presence of such outliers by using \textit{Mean Absolute Loss} (or L1 Loss) for training instead of the more popular \textit{Mean Squared Error} (or L2 Loss), it cannot be completely eliminated. 

\nomenclature{BDA}{Bloch Dispersion Analysis}
\begin{equation}
\label{eq:inn_modelling}
        f: \begin{bmatrix}
        \mathbf{x_m} \\ \mathbf{y_m} \\ \mathbf{x_f}  \\ \mathbf{y_f}  \\
        \end{bmatrix}
        \to  \begin{bmatrix}
            \boldsymbol{\omega}_{1}\\ \boldsymbol{\omega}_{2} \\ \boldsymbol{\omega}_{1}\\ \boldsymbol{\omega}_{2}
        \end{bmatrix}
\end{equation}

\begin{figure*}[]
\begin{subfigure}[]{.49\textwidth}
    \centering
    \includegraphics[width =\linewidth]{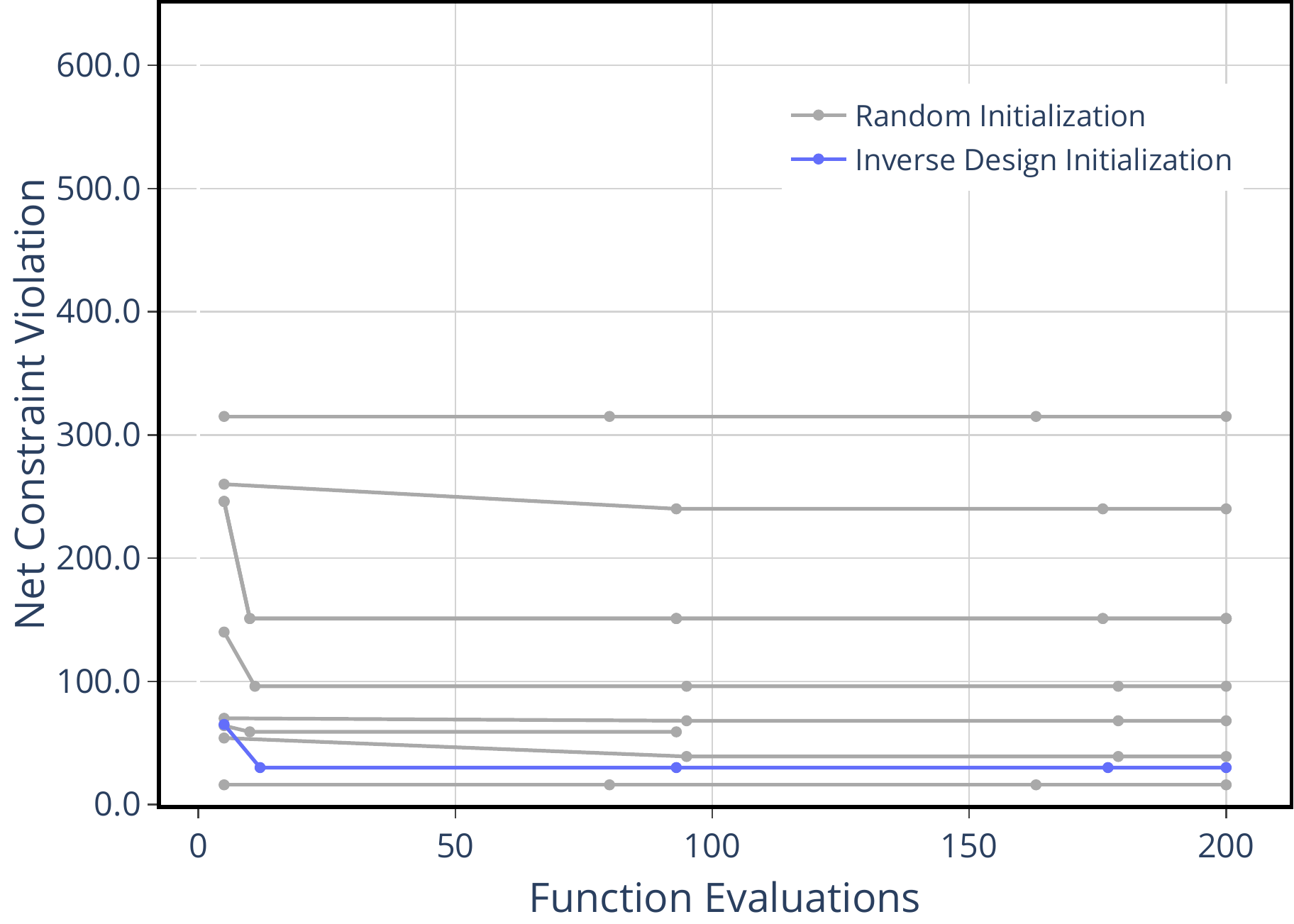}
    \caption{}
    \label{fig:convhist_1}
\end{subfigure} \hfill
\begin{subfigure}[]{.49\textwidth}
    \centering
    \includegraphics[width=\linewidth]{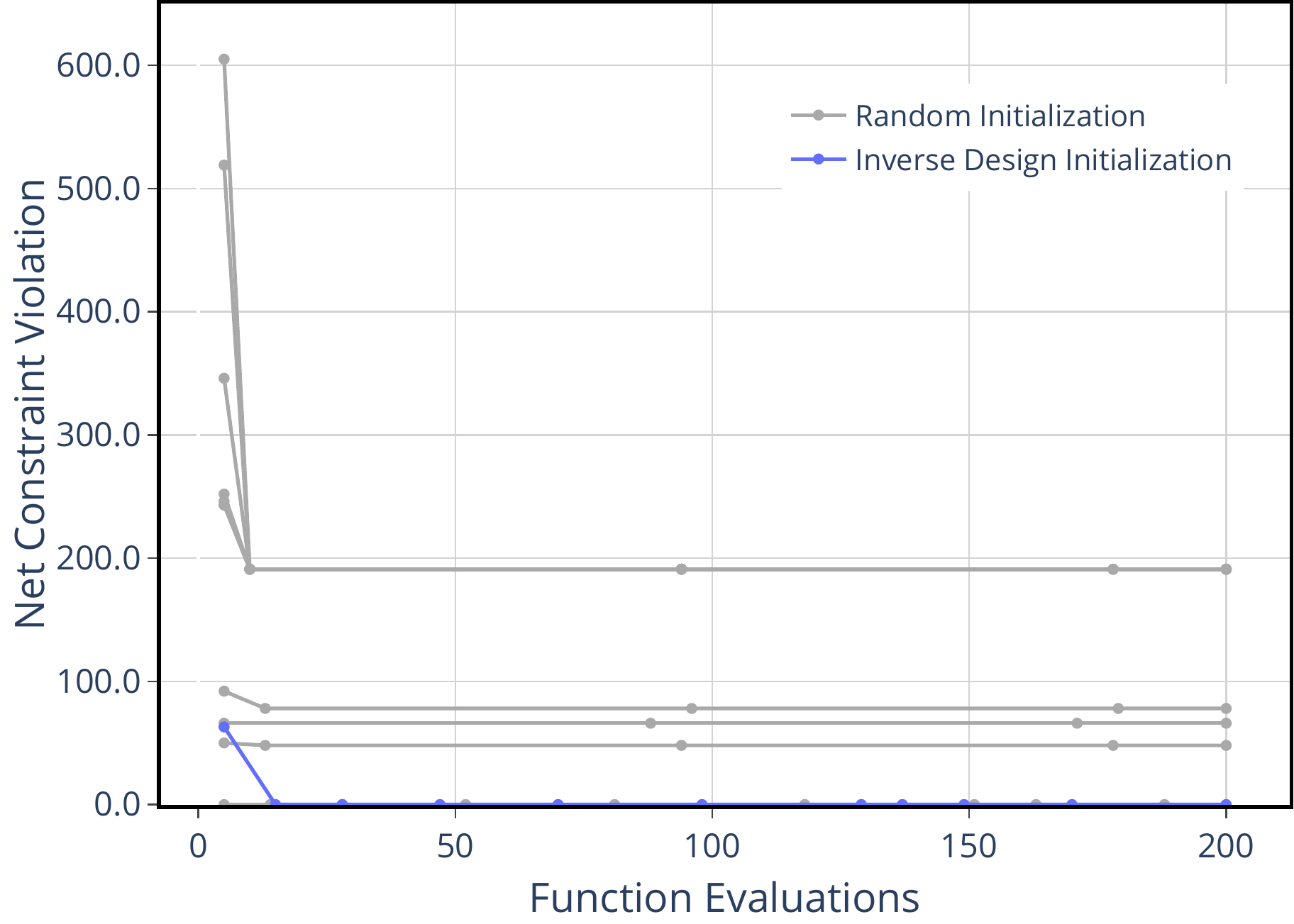}
    \caption{}
\label{fig:convhist_2}
\end{subfigure} \\ 
\begin{subfigure}[]{.49\textwidth}
    \centering
    \includegraphics[width =\linewidth]{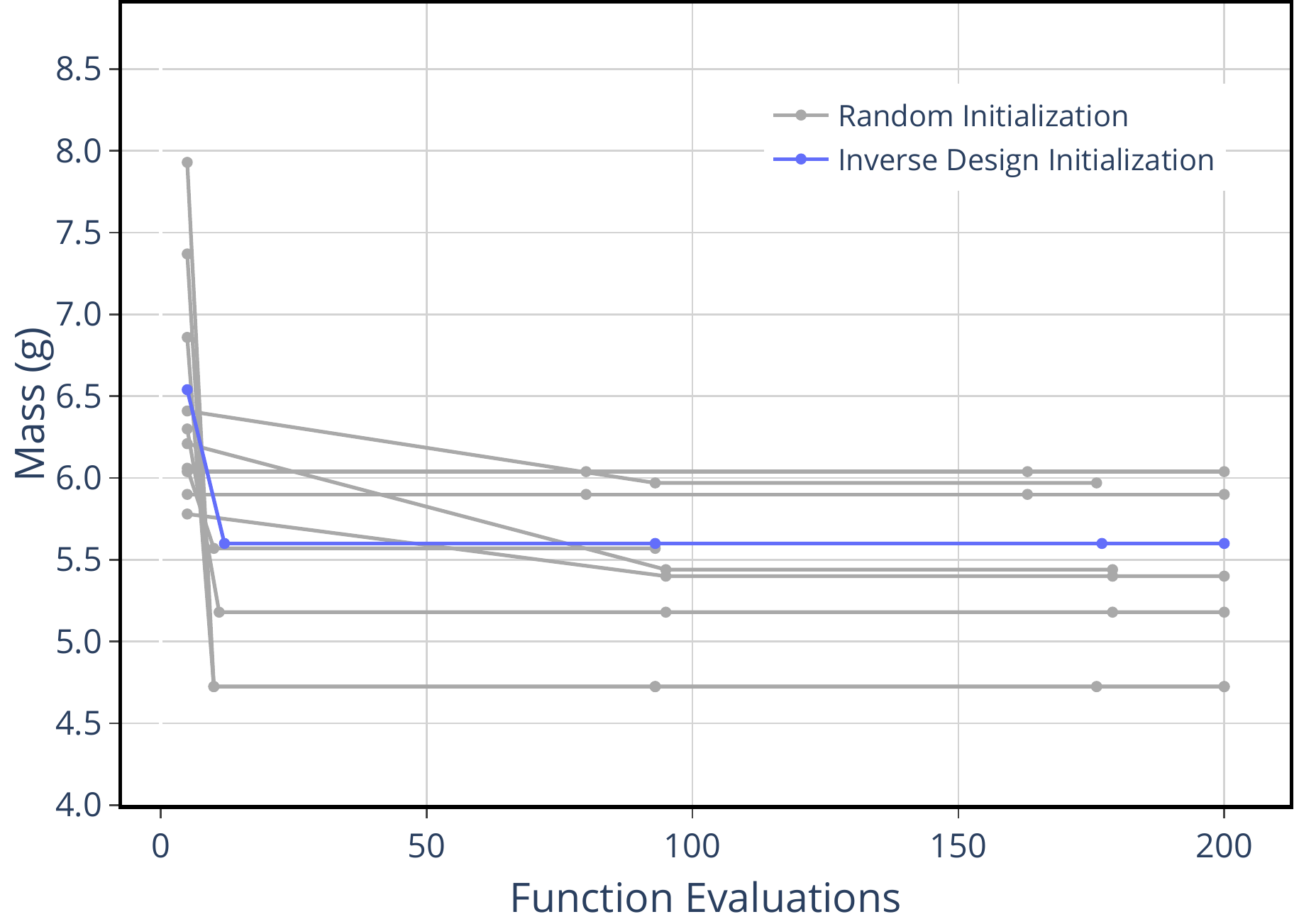}
    \caption{}
    \label{fig:masshist_1}
\end{subfigure} \hfill
\begin{subfigure}[]{.49\textwidth}
    \centering
    \includegraphics[width=\linewidth]{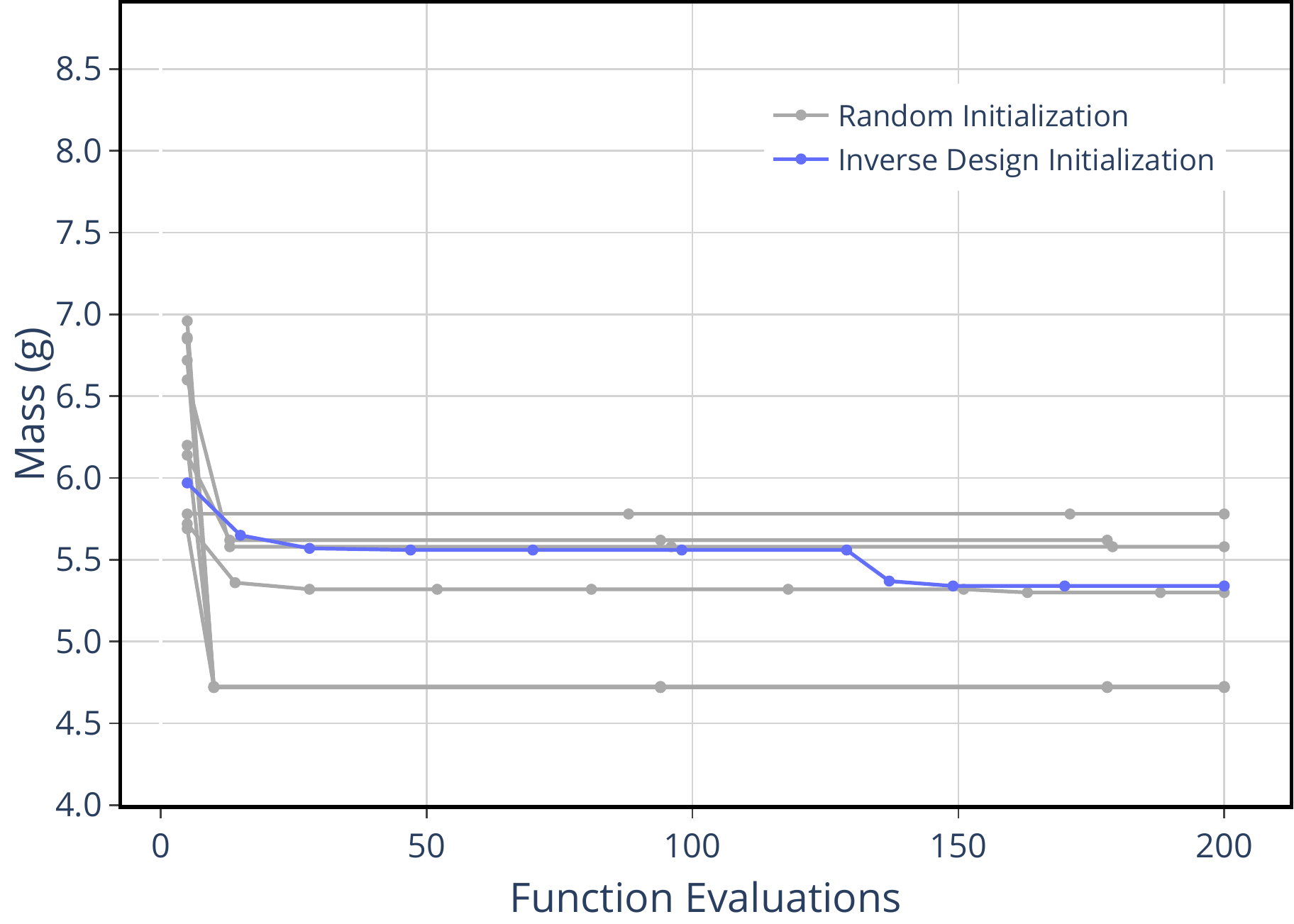}
    \caption{}
\label{fig:masshist_2}
\end{subfigure}
    \caption{Convergence history of SQP optimizations for the 1000-1700 Hz bandgap constraint: a) Constraint violation history with DNN; b) Constraint violation history with Dispersion Analysis (BDA); c) Objective function history with DNN; d) Objective function (Mass) history with BDA}
    \label{fig:const_hist_2}
\end{figure*}

\begin{table*}[]
\centering
\caption{Optimization results for inverse design of the 2D \lrems structure given \textit{Bandgap} query}
\footnotesize
\begin{tabularx}{\linewidth}{ >{\centering} X >{\centering}X >{\centering}X >{\centering}X >{\centering}X }
\toprule[0.12em]
\textbf{Bandgap Query \\ $\boldsymbol{\omega}_1$ to $\boldsymbol{\omega}_2$ Hz} & \textbf{Optimization Initialization*} & \textbf{ \# Trials} & 
 \textbf{Constraint Violation (Verified with BDA)} & \textbf{Mass (g)} \tabularnewline \midrule[0.12em]
 \multirow{4}{=} {1000-1700 Hz} 
                               & Random, Median run (High Fidelity) &  10 & 78 & 5.58 \tabularnewline 
                               & INN  (High Fidelity)               &  1  & 0 & 5.34 \tabularnewline \cline{2-5}
                               & Random, Median run (DNN based)     &  10 & 151 & 4.72   \tabularnewline
                               & INN    (DNN based)                 &  1  & 30  & 5.60      \tabularnewline \midrule[0.12em]
\multirow{4}{=} {800-1050 Hz}  
                               & Random, Median run (High Fidelity) &  10 & 446 & 6.02 \tabularnewline 
                               & INN (High Fidelity)                &  1  & 435 & 5.60  \tabularnewline \cline{2-5}
                               & Random, Median run (DNN based)     &  10 & 292 & 5.79 \tabularnewline 
                               & INN (DNN based)                    &  1  & 194 & 5.66 \tabularnewline \midrule[0.12em]   

\end{tabularx}

\label{tab:optimization}
\begin{flushleft}\footnotesize{*Random, best: The best results (in terms of feasibility, then mass) out of the 10 optimization runs that use random initialization}\end{flushleft}
\end{table*}
\subsection{Optimization Results}
\label{subsec:invdesign_perf}
We perform two different case studies, corresponding to two different user-desired bandgaps: \textbf{1)} bandgap at 1000--1700 Hz, and \textbf{2)} bandgap at 800--1050 Hz. The results of the first case study are shown here, with the latter shown in \ref{appendix:add_res}. For the given user-defined bandgaps, we use the INN executed in reverse to retrieve the design that most closely satisfies this bandgap property. This INN-retrieved design is then used as the initial design in the constrained forward optimization (solving Eq. \ref{eq:objfunc}) to find a better inverse design with minimum mass while still satisfying the bandgap property desired by the user. We also run 10 optimizations with randomly generated initial designs. The results of our optimizations (with the constraints estimated by the high-fidelity dispersion analysis) are listed in Table \ref{tab:optimization}, which compares the median scenario among the randomly initialized optimizations with the corresponding INN initialized optimizations. The two scenarios are such chosen so as to represent a larger bandgap  in a region where existence of bandgap is more likely (in the first case), and in the second case looking for the bandgap that is small but in a region where it's physically difficult to impose one. 
From Table \ref{tab:optimization} it can be seen that the INN initialized optimizations perform better than the median randomly initialized optimizations, which is attributed to the INN's ability to find a feasible or closer to feasible inverse solution to start with, thereby likely decreasing the computational cost of optimization or improving the optimization performance. This observation is further corroborated by the convergence histories of the optimization shown in Fig. \ref{fig:const_hist_2}.  
In this figure, the outcomes of the outcomes of the surrogate based optimization (using DNN) are shown on the left and the high-fidelity optimization (using BDA) are shown on the right. The top plots show the variation of net constraint violation (violation of the specified bandgap) with function evaluations, and the bottom plots show the corresponding history of the objective function (i.e., mass). As seen from Figs. \ref{fig:convhist_1} and \ref{fig:const_hist_2}, given the specified bandgap, the INN is able to retrieve an initial design that in general has a smaller violation of the bandgap constraint compared to random initializations, giving it a greater likelihood of finding closer-to-feasible design with optimization that use the same number of maximum function evaluations as the randomly initialized runs.

When using INN for inverse retrieval and then performing forward optimization with DNN, the total computing time is measured to be in the order of 10 minutes, running on a 6 core Intel i7-9750H and a Nvidia RTX 2060 (with the DNN running on the GPU). 
Such levels of computational efficiency clearly points to the ability of the framework (post training of the ML models) to serve in the role of an on-demand inverse design process.


\begin{figure}[]
    \centering
    \includegraphics[width=\linewidth]{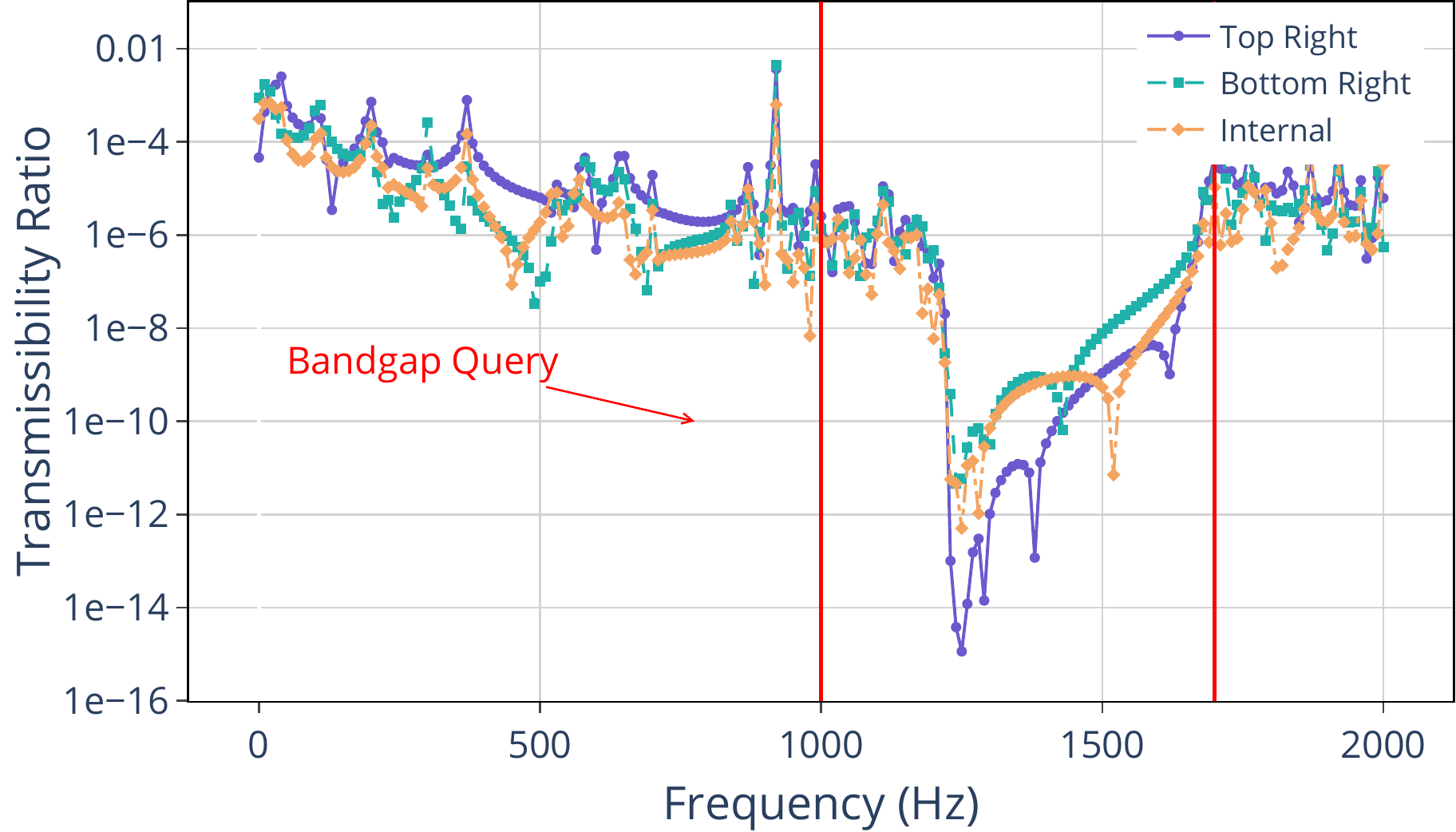}
\label{fig:tf_inn}\vspace{-0.5cm}
    \caption{Transmissibility Ratio of optimized design given by INN-initialized DNN-based SQP -- 1000-1700 Hz case}
    \label{fig:tf}
    \vspace{-0.5cm}
\end{figure}

\begin{figure*}[]
\begin{subfigure}[]{.4\textwidth}
    \centering
    \includegraphics[width =0.7\linewidth]{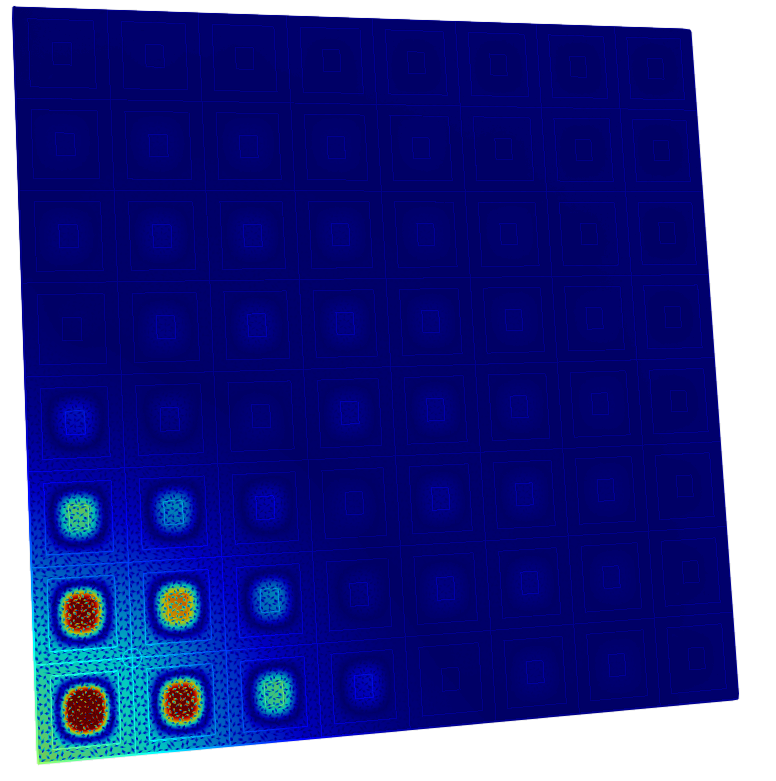}
    \caption{}

\end{subfigure}
\begin{subfigure}[]{.4\textwidth}
    \centering
    \includegraphics[width=0.7\linewidth]{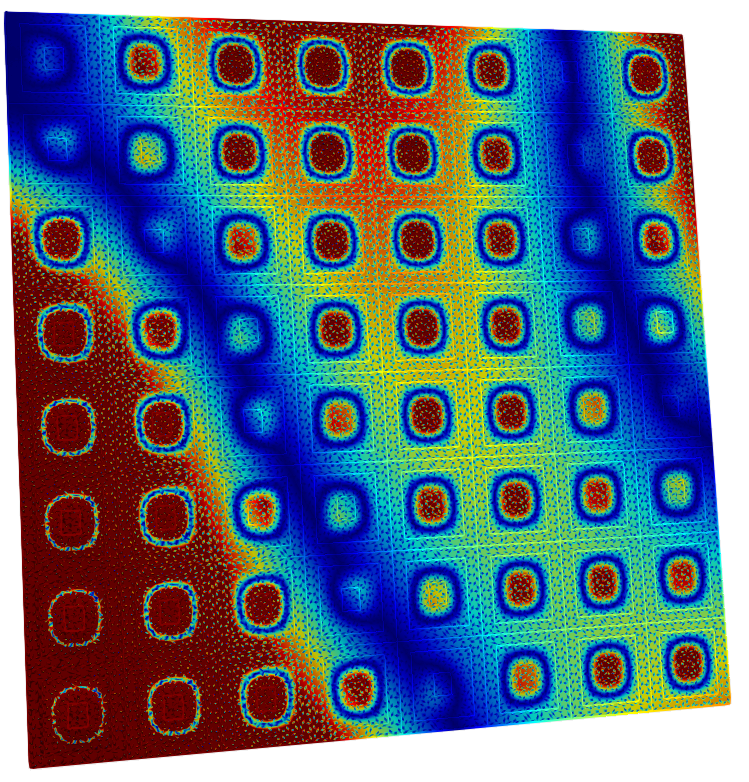}
    \caption{}

\end{subfigure}
\begin{subfigure}[]{.15\textwidth}
    \centering
    \includegraphics[width=0.7\linewidth]{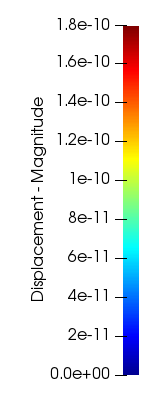}

\end{subfigure}
    \caption{FEA based verification: Representative example of displacement plots showing vibration transmission a) Within Bandgap: 1020 Hz; b) Outside Bandgap: 1120 Hz}
    \label{fig:inn_disp}
\end{figure*}

Figure \ref{fig:tf} shows the transmissibility ratio of the finite structure comprising $8\times 8$ unit cells of the optimum design given by the INN intialized optimizations performed using DNN. 
From this figure it is apparent that the actual behaviour of the finite structure is very close to the behaviour predicted by the INN. Furthermore, Fig. \ref{fig:inn_disp} shows a representative example of the displacement plots within and out of the bandgap, clearly demonstrating that for a representative excitation at a frequency within the bandgap is readily damped by only a few layers of unit cells, while excitations at a frequency outside the desired bandgap continue to persist in the system. 

\section{Conclusion}
\label{sec:Conclusion}

In this paper we develop an inverse design framework to provide on-demand design of 2D LREM structures. The framework involved generating an initial set of samples using ``Bloch Dispersion Analysis'' and then filtering the frequency points obtained in order to get the bandgaps. Then, an Invertible Neural Network or INN and a traditional Deep Neural Network were trained on the same set of samples and their modeling performance was compared on unseen samples to validate acceptable levels of accuracy for further usage. Here, both the NN models predict the bandgap bounds given the geometry of the unit cell, defined in terms of the width of the outer matrix and middle soft filler layers. 
We then used the INN to generate an initial design corresponding to user specified bandgap, resulting it a close-to-feasible inverse design, which is used to initialize the forward constrained optimization for finding better inverse designs that also minimize mass. This approach was observed to in general provide better designs in terms of satisfying the specified bandgap constraints and mass, compared to randomly initialized optimizations. 
Finally, the optimization results were verified by performing a finite element harmonic analysis on the entire finite structure containing 64 unit cells and analyzing it's transmissibility ratio w.r.t the excitation frequency. The FEA results show that the finite structure has enough number of unitcells to achieve the bandgap behaviour predicted by the dispersion analysis. 

A fundamental issue with modeling bandgaps is the non-existence of one (in a given global frequency range) in many training samples, which was dealt with in a simplified manner by assigning out-of-range bandgap bounds; such an approach can however be detrimental to the overall accuracy of the INN, and better ways of dealing with issue is an immediate direction of future work. There's also the issue of feasible solutions not existing for any randomly specified user-requested bandgap, in which case the INN becomes forced to predict a design for a bandgap that may not be physically meaningful within the given material and geometric settings. Future work could therefore also look at approaches to filter out or prohibit such requests, by learning from the initial training data-set the likelihood of bandgaps existing within the current range of design parameters. Lastly, while the current framework has been applied to design unit cells, its applicability to design functional systems comprising 2D metamaterials remains to be explored, in order to fully demonstrate the potentials of fast on-demand retrieval of inverse designs. 

\bibliographystyle{IEEEtran}
\bibliography{manaswin}
\appendix
\begin{figure}[]
    \centering
    \includegraphics[width=\linewidth]{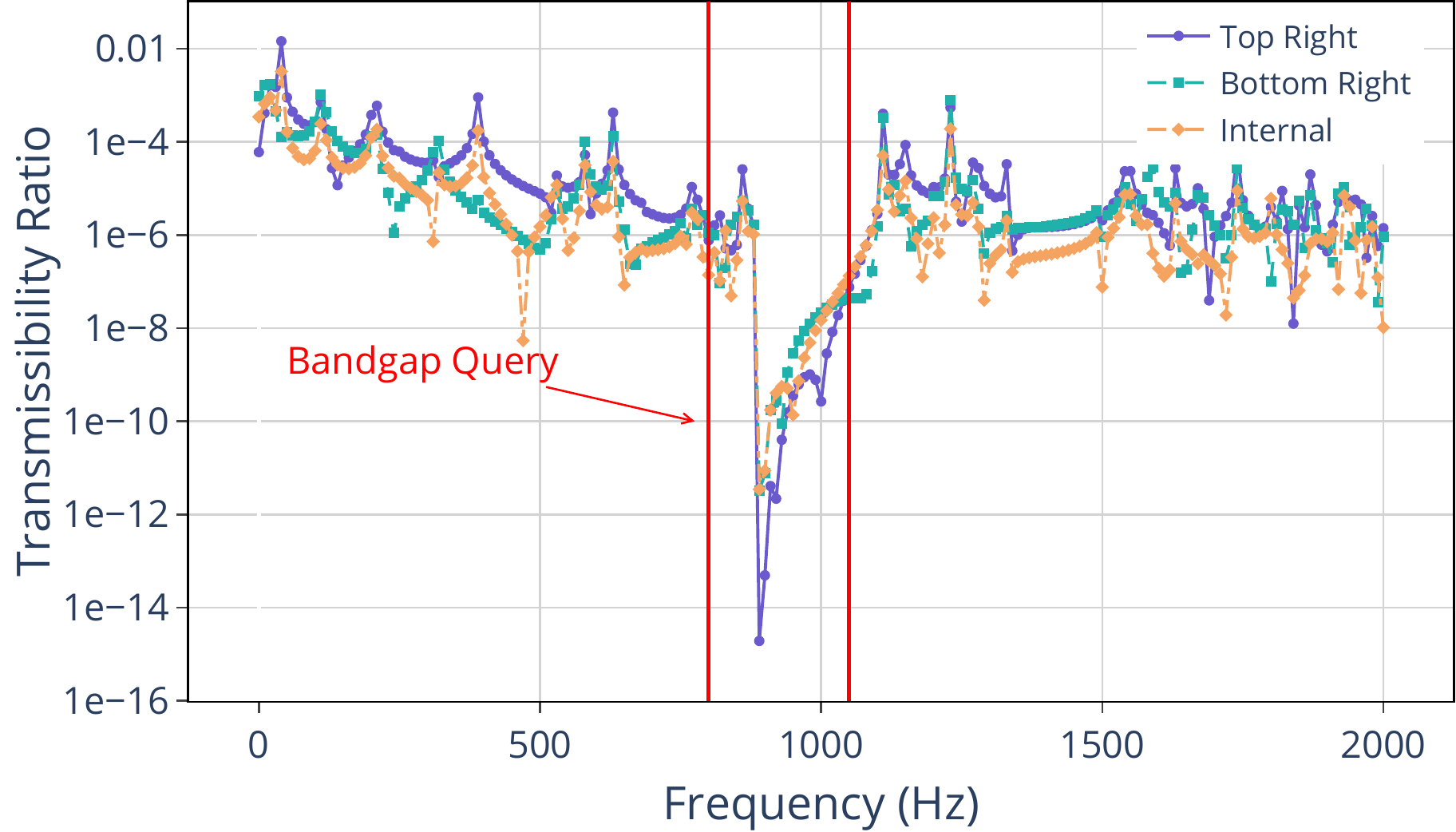}
\label{fig:tf_inn_2}
    \caption{Transmissibility Ratio of the result of DNN based SQP Optimization with 800-1050 Hz Constraint INN Initialization}
    \label{fig:tf_2}
\end{figure}

\begin{figure}[t]
\begin{subfigure}[]{.49\textwidth}
    \centering
    \includegraphics[width =0.8\linewidth]{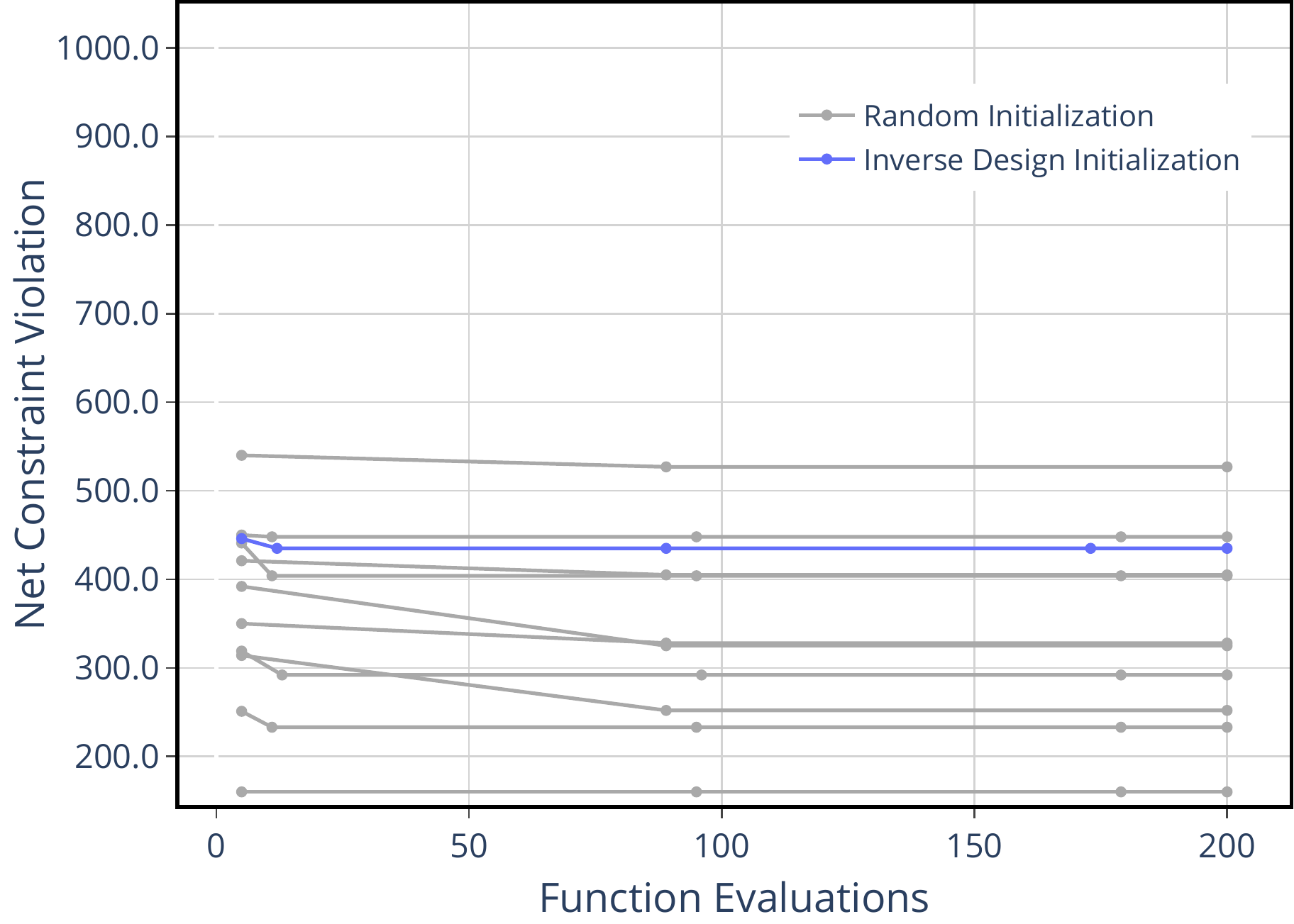}
    \caption{}
\end{subfigure}\hfill
\begin{subfigure}[]{.49\textwidth}
    \centering
    \includegraphics[width=0.8\linewidth]{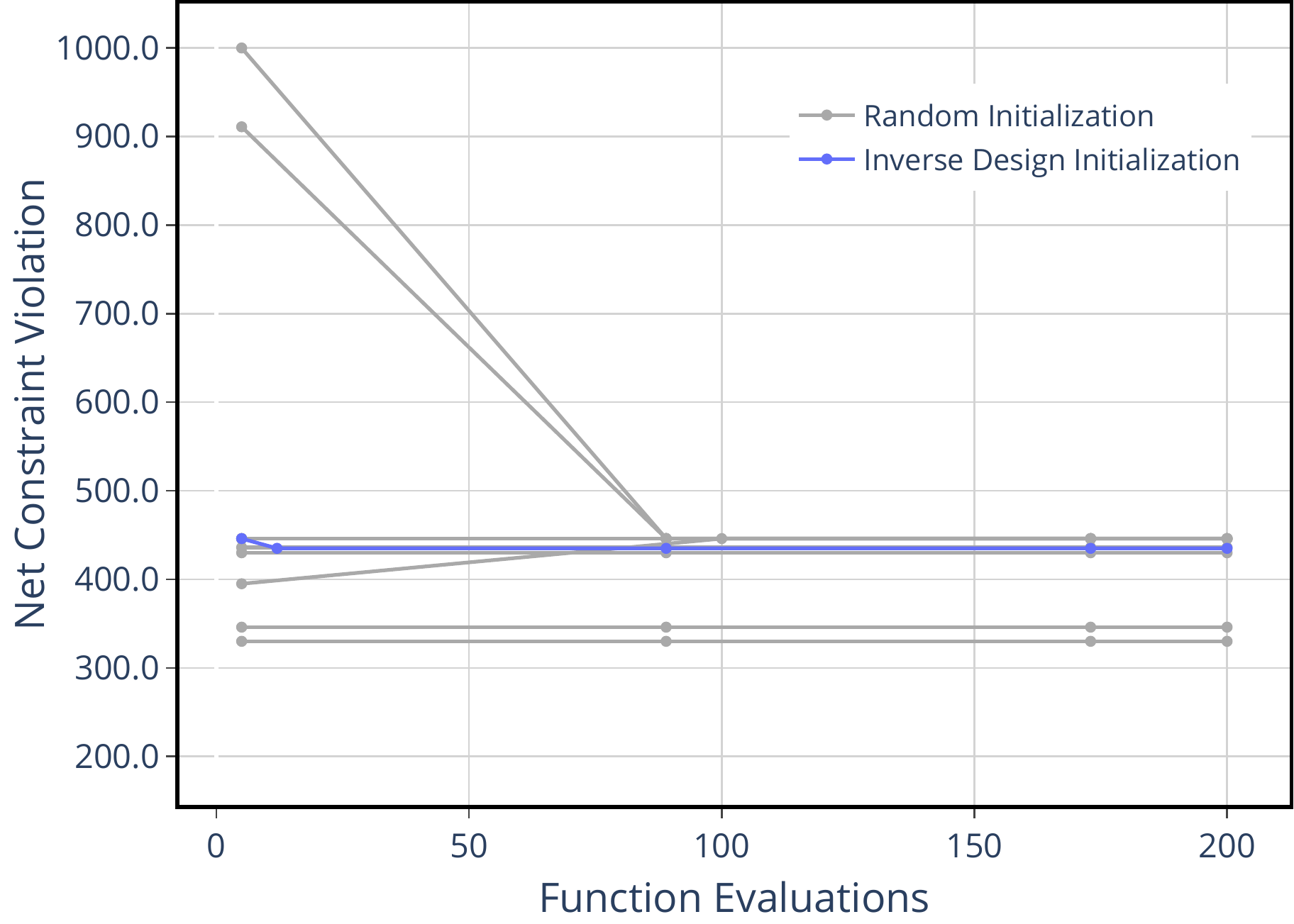}
    \caption{}
\end{subfigure}
    \caption{Convergence history of SQP optimizations with 800-1050 Hz bandgap constraint a) Constraint history using DNN as surrogate; b) Constraint history using High Fidelity Dispersion Analysis }
    \label{fig:const_hist_1}
\end{figure}
\renewcommand{\thesection}{Appendix \Alph{section}}
\section{Invertible Neural Networks}

\begin{figure}[]
\begin{subfigure}[]{.49\textwidth}
    \centering
    \includegraphics[width=0.8\linewidth]{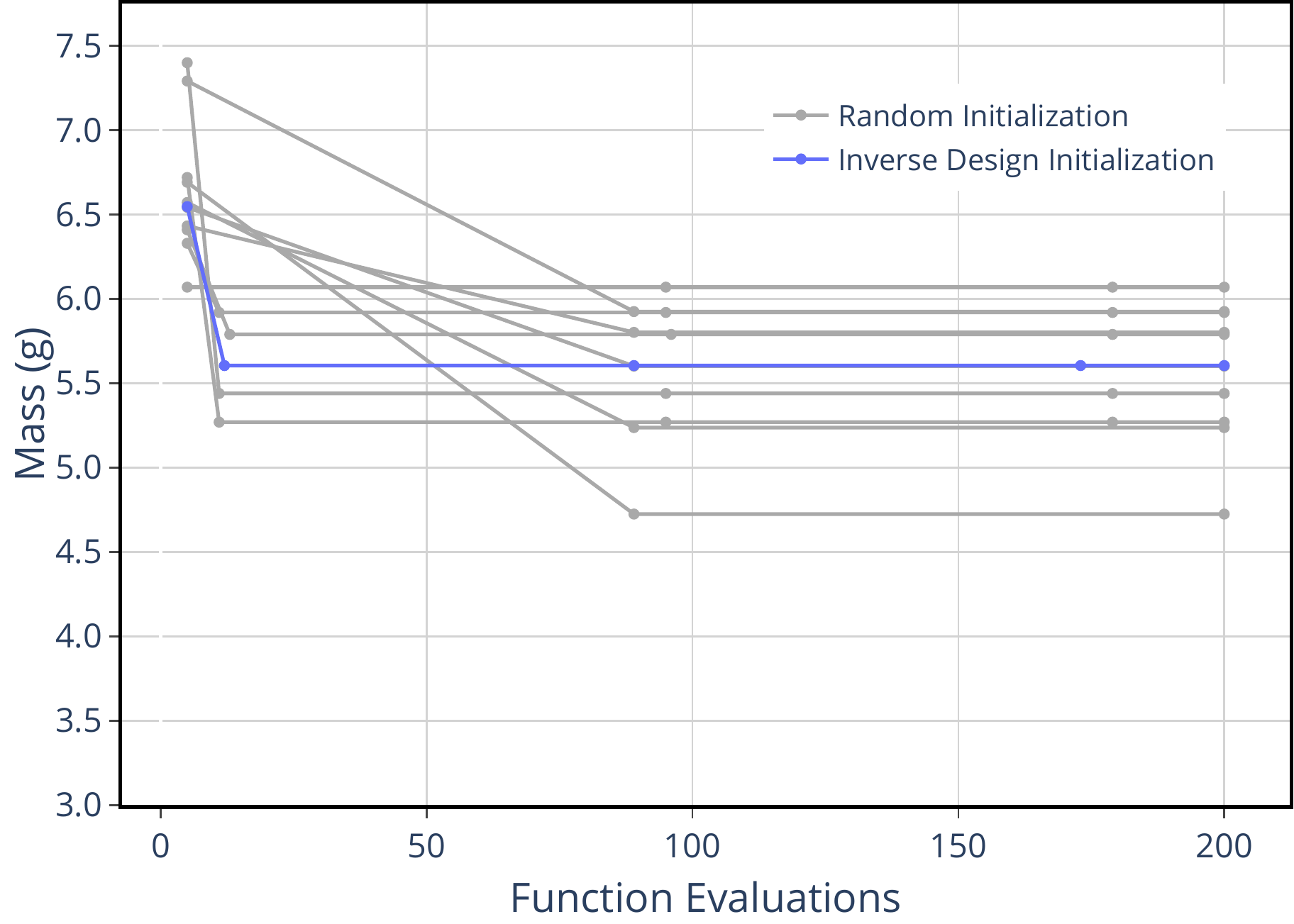}
    \caption{}
\end{subfigure} \hfill
\begin{subfigure}[]{.49\textwidth}
    \centering
    \includegraphics[width=0.8\linewidth]{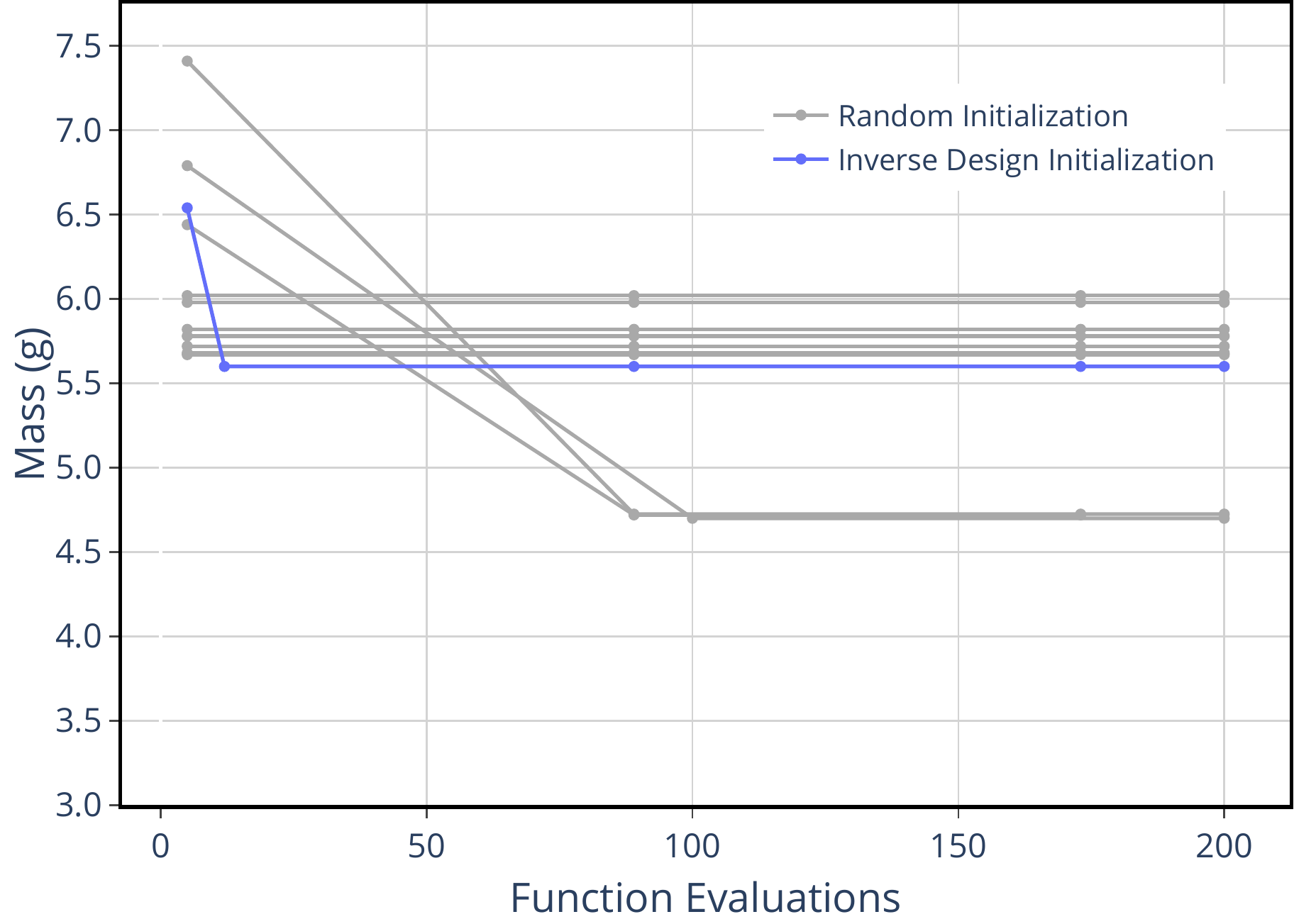}
    \caption{}
\end{subfigure}
    \caption{Convergence history of SQP optimizations with 800-1050 Hz bandgap constraint a) Objective Function(Mass) history using DNN as Surrogate; b) Objective Function(Mass) using High Fidelity Dispersion Analysis}
    \label{fig:const_hist_4}
\end{figure}

\label{appendix:INN}
\begin{figure*}[ht]
\begin{subfigure}
[]{.65\textwidth}
    \includegraphics[width = \linewidth]{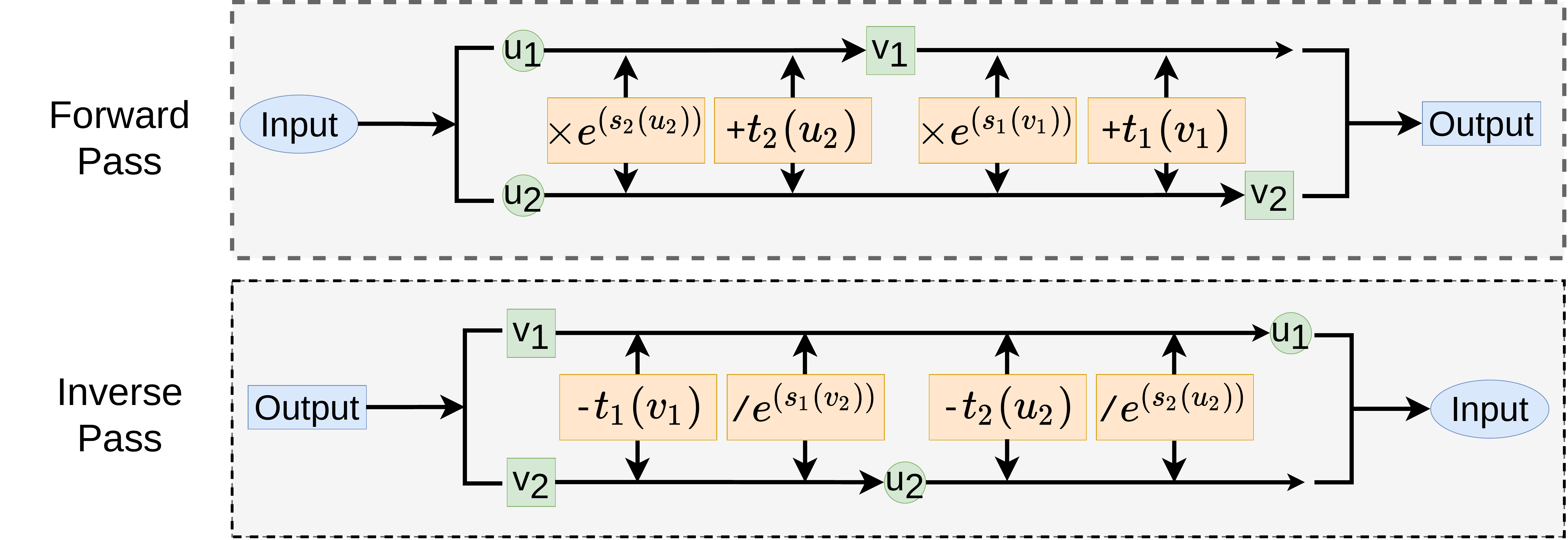}
    \caption{}
    \label{fig:Invblock}
\end{subfigure}
\begin{subfigure}[]{.34\textwidth}
    \centering
    \includegraphics[width =\textwidth]{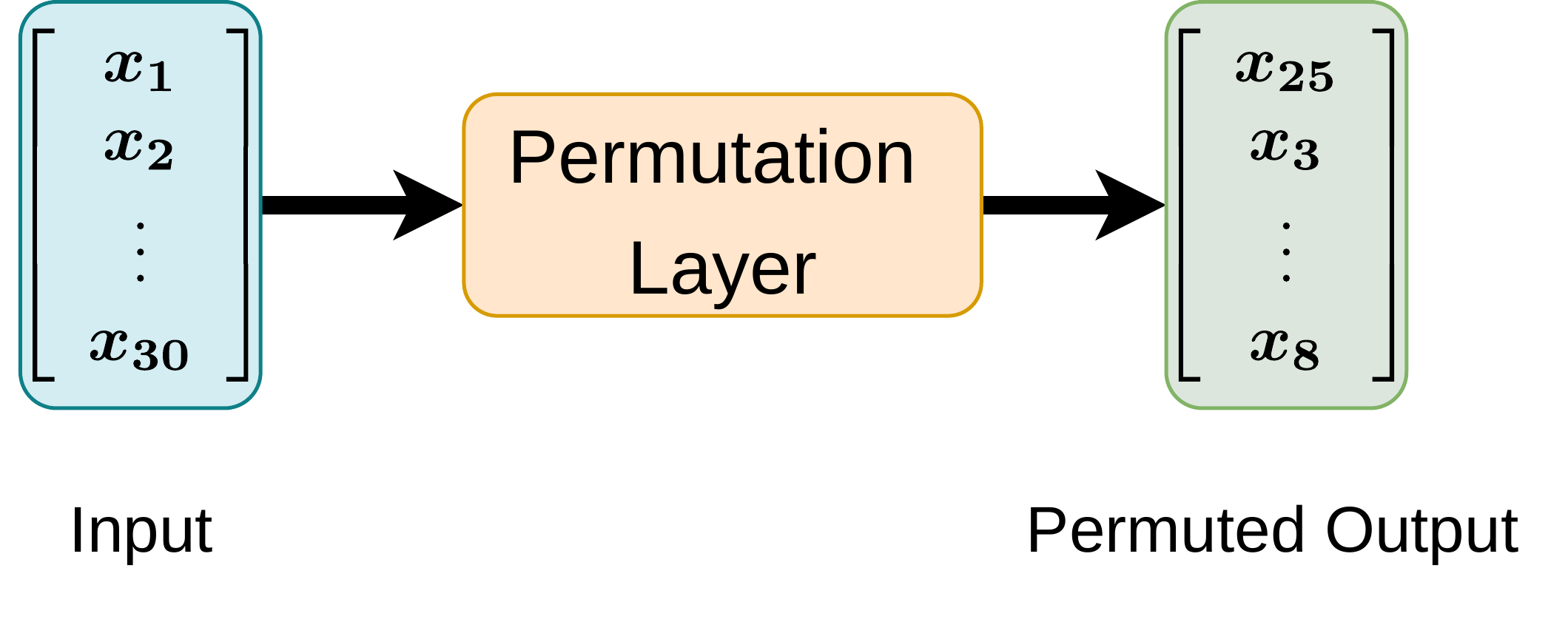}
    \caption{}
    \label{fig:perm_block}
\end{subfigure}\\
\begin{subfigure}[]{\textwidth}
    \centering
    \includegraphics[width = \linewidth]{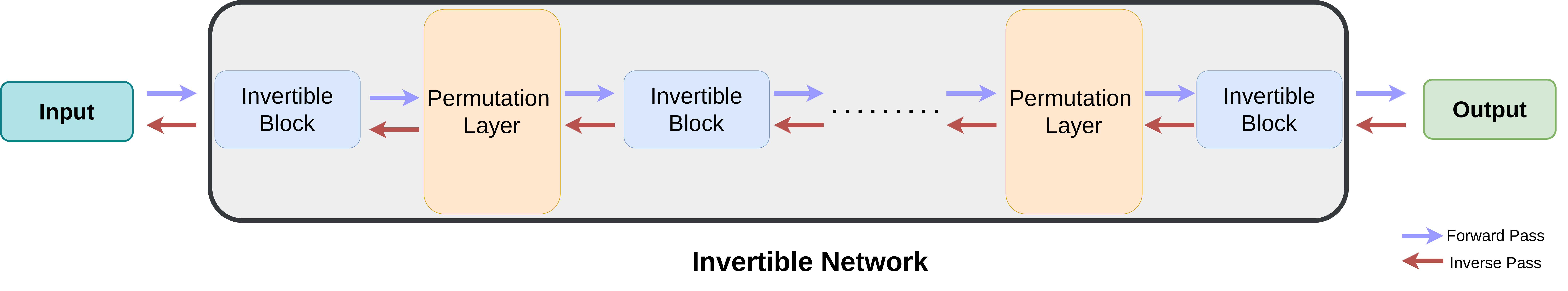}
    \caption{}
    \label{fig:INN_framework}
\end{subfigure}
\caption{Invertible Neural Network (INN) architecture: a) Invertible blocks; b) Permutation block; c) Entire INN}
\label{fig:innarch}
\end{figure*}
\section{Additional Results of LREM Inverse Design for Bandgap at 800--1050 Hz}
\label{appendix:add_res}


Figure \ref{fig:innarch} shows the architecture of the INN. 
The structure, which is shown in Fig. \ref{fig:innarch} top, constitutes one INN block. The block is a deep network formed by a series of such blocks with permutation layers in between to shuffle the input to the next block pseudo-randomly.
In the above architecture, it is assumed that the inputs and outputs of the model are of the \textbf{same size}. If they are not of the same size, then:
\begin{enumerate}
    \item If the size of outputs \textgreater size of inputs, another set of variables referred to as the \textit{Latent variables} are defined, the size of which is equal to the difference of the input and output sizes.
    \item If the size of inputs \textless size of outputs, the inputs are padded with zeros.
\end{enumerate} 

Latent variables are used in the INN to compensate for the information loss associated with dimension reduction. They are especially useful in scenarios where multiple inputs map to a single output, i.e., in cases where the inverse problem is \textit{ill-posed}. When running the network in reverse, the input variables' distribution for the given output can be obtained by varying the latent variables with the same set of outputs.
The \INN uses the relationship between these latent variables and the actual parameters in order to capture the input distribution for a given output

\end{document}